\documentstyle[12pt,epsf,epsfig,amsmath,amsfonts]{article}

\def\be{\begin{equation}}
\def\ee{\end{equation}}
\def\ba{\begin{eqnarray}}
\def\ea{\end{eqnarray}}

\def\bdm{\begin{displaymath}}
\def\edm{\end{displaymath}}

\def\ga{~\mbox{\raisebox{-.6ex}{$\stackrel{>}{\sim}$}}~}
\def\bq{\begin{quote}}
\def\eq{\end{quote}}

 at 10truept

\newcommand{\beq}{\begin{equation}}
\newcommand{\eeq}{\end{equation}}
\newcommand{\bea}{\begin{eqnarray}}
\newcommand{\eea}{\end{eqnarray}}
\newcommand{\beqa}{\begin{eqnarray}}
\newcommand{\eeqa}{\end{eqnarray}}

\def\ga{~\mbox{\raisebox{-.6ex}{$\stackrel{>}{\sim}$}}~}

\def\ltap{\ \raise.3ex\hbox{$<$\kern-.75em\lower1ex\hbox{$\sim$}}\ }
\def\gtap{\ \raise.3ex\hbox{$>$\kern-.75em\lower1ex\hbox{$\sim$}}\ }
\def\gl{\ \raise.5ex\hbox{$>$}\kern-.8em\lower.5ex\hbox{$<$}\ }
\def\roughly#1{\raise.3ex\hbox{$#1$\kern-.75em\lower1ex\hbox{$\sim$}}}

\begin{document}

\thispagestyle{empty}
\begin{flushright}
September 2013
\end{flushright}
\vspace*{1.25cm}
\begin{center}
{\Large \bf Sequestering the Standard Model Vacuum Energy}\\

\vspace*{1.6cm} {\large Nemanja Kaloper$^{a, }$\footnote{\tt
kaloper@physics.ucdavis.edu} and Antonio Padilla$^{b, }$\footnote{\tt
antonio.padilla@nottingham.ac.uk} }\\
\vspace{.5cm} {\em $^a$Department of Physics, University of
California, Davis, CA 95616, USA}\\
\vspace{.5cm} {\em $^b$School of Physics and Astronomy, 
University of Nottingham, Nottingham NG7 2RD, UK}\\

\vspace{1.5cm} ABSTRACT
\end{center}
We propose a very simple reformulation of General Relativity, which completely sequesters from gravity {\it all} of the vacuum energy 
from a matter sector, including all  loop corrections and renders all contributions from phase transitions automatically small. The idea 
is to make the dimensional parameters in the matter sector functionals of the $4$-volume element of the universe. For them to be nonzero, the universe 
should be finite in spacetime. If this matter is the Standard Model of particle physics, our mechanism prevents any of its vacuum energy, 
classical or quantum, from sourcing the curvature of the universe. The mechanism is consistent with the large hierarchy between the 
Planck scale, electroweak scale and curvature scale, and early universe cosmology, including inflation. Consequences of our proposal 
are that the vacuum curvature of an old and large universe is not zero, but very small, that $w_{DE} \simeq -1$ is a transient, and that 
the universe will collapse in the future.

\vfill \setcounter{page}{0} \setcounter{footnote}{0}
\newpage

The cosmological constant problem is the most severe naturalness
problem in fundamental physics \cite{zeldovich,wilczek,wein}. It follows from the
Equivalence Principle of General Relativity (GR) which
asserts that {\it all} forms of energy curve spacetime. So,
even the energy density of the vacuum, which contributes 
to the cosmological constant, sources the curvature of the spacetime,
generically giving it huge contributions.
One can add a classical piece to the cosmological constant
and tune it with tremendous precision to
cancel the vacuum energy. 
However, this tuning is unstable: any change of the matter sector 
parameters or addition of loop corrections to vacuum energy 
dramatically shifts the value of vacuum energy, by ${\cal O}(1)$ in the units of the UV cutoff. 
To neutralize it one must retune the classical term by hand order by order in perturbation 
theory\footnote{If exact, SUSY and/or conformal symmetry can enforce the vanishing of vacuum
energy. In the real world are broken, which induces vacuum energy given 
by the fourth power of the breaking scale \cite{wein, dreitlein,linde,veltman}.}.

In this {\it Letter} we present a mechanism which provides a remedy,Ê
ensuring that {\it all} the vacuum energy
%nk is
from a matter sectorÊ is 
sequestered from gravity. ThisÊ includes matter  loop corrections (not involving virtual gravitons) which 
are invisible to gravity, and Êcontributions from phase
transitions, which are automatically small at late times.
Our idea is to make all scales in this 
matter sector functionals of the $4$-volume element of the universe. For the scales to be nonzero, the universe 
should be finite in spacetime, collapsing in the future. If the matter sector is the Standard Model of particle physics,
our mechanism prevents it from generating large contributions to the net cosmological constant, and therefore to the
curvature of the background universe. The mechanism is a very minimal modification of General Relativity, without any new propagating degrees of freedom. We formulate it adding to the action  auxiliary fields with an extra term, which is {\it not} integrated over and is completely covariant. This  
subtracts `historic averages' of the matter stress energy from the gravitational sources, and removes the vacuum energy contributions from the field equations. Nonetheless, there is still an effective net nonzero cosmological term, but now {\it i)} it is purely classical, set by the complete evolution of the geometry, {\it ii)} it is a `cosmic average' of the values of non-constant sources, and so {\it iii)} it is automatically small in universes which grow 
large and old\footnote{This doesn't automatically make it fit the observational 
data, but it makes the tunings needed far gentler.}. 
In the limit of (semi) classical gravity there 
are absolutely no dynamical pathologies. All the propagating degrees of freedom obey standard 
second order field equations compatible with
local Poincare symmetry and diffeomophism invariance, and the spectrum of fluctuations is the same as in conventional GR with 
minimally coupled matter.
The mechanism is consistent with phenomenological requirements, specifically with large hierarchies between the Planck scale,
electroweak scale and vacuum curvature scale, and with early universe cosmology including inflation.
Observable consequences of our proposal are 
that the vacuum curvature of an old and large universe is not zero, but very small, 
that $w_{DE} \simeq -1$ is a transient, 
and that the universe is compact in its spacetime extent, crunching in the future.

Our mechanism can be described by writing the action for a matter sector which couples to gravity 
as 
\be
S= \int d^4 x \sqrt{g} \left[ \frac{M^2_{Pl}}{2} R  - \Lambda - {\lambda^4} {\cal L}(\lambda^{-2} g^{\mu\nu} , \Phi) \right] +\sigma\left(\frac{ \Lambda}{\lambda^4 \mu^4}\right) 
 \, , ~~
\label{action}
\ee
where matter couples minimally to the rescaled metric $\tilde g_{\mu\nu}=\lambda^2 g_{\mu\nu}$.  The parameter $\lambda$ sets the hierarchy between the matter scales and the Planck scale, since $m_{phys}/M_{Pl} \propto \lambda m/M_{Pl}$,
 where $m_{phys}$ is the physical mass scale and $m$ is the bare mass in the Lagrangian. In conventional GR (or unimodular gravity) the variable $\Lambda$ would be an arbitrary classical contribution to the total cosmological constant. 
We treat  the parameters $\Lambda$ and $\lambda$ as dynamical variables without {\it any} local dynamics -- i.e. just as auxiliary fields. To define the field equations we vary (\ref{action}) with respect to $\Lambda$ and $\lambda$ in addition to other variables with local propagating modes, as in formulations of unimodular gravity \cite{finkel,buch,henteitel,unruh}. Here,  in contrast to old approaches, we add the function 
$\sigma$ {\it outside} of the integral, to fix the matter scales as functionals of $\int d^4x \sqrt{g}$. 
The external function $\sigma$ is an odd (to allow for solutions with vacuum energy of either sign for $\lambda > 0$) differentiable function, to be determined by phenomenology. 
In QFT,  it would completely drop out of the calculation of any observables, but
with gravity turned on, it affects the dynamics of the metric determinant  $g = -\det(g_{\mu\nu})$ sector. The scale $\mu$ is also chosen phenomenologically. 

From (\ref{action}) one can see that all vacuum energy contributions coming from
the Lagrangian $\sqrt{g} \lambda^4 {\cal L}(\lambda^{-2} g^{\mu\nu} , \Phi)$, must scale with $\lambda$ as $\lambda^4$,
even after the logarithmic corrections are included, provided that a regulator of the QFT is defined to also couple minimally to $\tilde g_{\mu\nu}$\footnote{We can take an appropriate system of Pauli-Villars regulator fields for ${\cal L}$ and directly couple them to $\tilde g_{\mu\nu}=\lambda^2 g_{\mu\nu}$. That ensures the cancellation of $\lambda$ in loop logarithms.}.
This follows from diffeomorphism invariance of the theory, which guarantees that the full effective Lagrangian computed from 
$\sqrt{g} \lambda^4 {\cal L}(\lambda^{-2} g^{\mu\nu} , \Phi)=\sqrt{\tilde g} {\cal L}(\tilde g^{\mu\nu} , \Phi)$, 
including all quantum corrections, still couples to the exact same $\tilde g_{\mu\nu}$ \cite{selft}. 
For our purposes in this {\it Letter} it suffices to consider gravity 
as a purely (semi) classical theory, and focus on the quantum effects from matter alone\footnote{In this limit the Weinberg's no-go theorem \cite{wein} governs the (lack of) adjustment of vacuum energy.}. 
After canonically normalizing the matter fields in ${\cal L}$, the matter mass scales that enter in physical observables scale as
$m_{phys} \propto \lambda m$, where $m$ are `bare' parameters in ${\cal L}$. So the vacuum energy, including all loop contributions to ${\cal L}_{eff}$, scales as $V_{vac} = \lambda^4 \langle 0 | {\cal L}_{eff} | 0 \rangle$.

The field equations that follow from varying the action (\ref{action}) with respect to (the `constants') $\Lambda, \lambda$ are
\be
\frac{\sigma'}{\lambda^4\mu^4} = \int d^4x \sqrt{g} \, , ~~~~~~~~~~~~~~~~ 4\Lambda \frac{ \sigma' }{\lambda^4\mu^4} 
= \int d^4x \sqrt{g} \, \lambda^4 \, \tilde T^{\mu}{}_\mu \, ,
\label{varsl}
\ee
where $\tilde T_{\mu\nu}=-\frac{2}{\sqrt{\tilde g}} \frac{\delta S_m}{\delta \tilde g^{\mu\nu}}$ is the energy-momentum tensor defined in the `Jordan frame'. To rewrite it in the `physical' frame, in which matter sector is canonically normalized, we note that 
 $T^\mu{}_\nu=\lambda^4 \tilde T^\mu{}_\nu$. Here $\sigma' = \frac{d\sigma(z)}{dz}$, and as long as it is nonzero\footnote{And non-degenerate: it can't be the pure logarithm, since then Eqs. (\ref{varsl}) turn into two independent constraints.}, we can eliminate it from the two Eqs. (\ref{varsl}) to find $\Lambda = \frac14 \langle T^\mu{}_\mu \rangle$, where we defined the $4$-volume average of $Q$ by
%nk the $4$-volume average of a quantity by 
$\langle Q\rangle ={\int d^4 x \sqrt{g} \, Q}/{\int d^4 x \sqrt{g}}$.
 
The variation of (\ref{action}) with respect to $g_{\mu\nu}$ yields $M_{Pl}^2 G^\mu{}_\nu = -\Lambda \delta^\mu{}_\nu + \lambda^4 \tilde T^\mu{}_\nu$,
which, by eliminating $\Lambda$ and canonically normalizing the matter sector, becomes
\be
M_{Pl}^2 G^\mu{}_\nu= T^\mu{}_\nu-\frac{1}{4} \delta^\mu{}_\nu \langle T^\alpha{}_\alpha \rangle \, ,
\label{eeqs}
\ee
where $G^{\mu}{}_\nu$ is the standard Einstein tensor. Eq. (\ref{eeqs}) is the key: it is the full system of {\it ten} field equations, with 
the trace equation {\it included}, and with the trace of the 4-volume historic average of the stress energy tensor of matter 
%nk is 
subtracted from the rhs!
This is unlike unimodular gravity \cite{finkel,buch,henteitel,unruh}, where although the restricted variation removes the trace equation that involves the vacuum energy, it comes back along with an arbitrary integration constant, after using the Bianchi identity. Here
there are {\it no} hidden equations nor integration constants, all the sources are automatically accounted for in (\ref{eeqs}).

Hence the hard cosmological constant, be it a classical contribution to ${\cal L}$ in (\ref{action}), or quantum vacuum correction calculated to any order in the loop expansion, never contributes to the field equations (\ref{eeqs}). Indeed, if we write 
${\cal L} = \Lambda_0 + V_{vac} + {\cal L}_{local}$, by our definition of the historic average, 
$\langle \Lambda_0 + V_{vac} \rangle \equiv \Lambda_0 + V_{vac}$. Next defining 
$\tau_{\mu\nu}= \frac{2}{\sqrt{g}} \frac{\delta}{\delta g^{\mu\nu}} \int d^4x \sqrt{g} \lambda^4 {\cal L}_{local}(\lambda^{-2} g^{\mu\nu}, \Phi)$
we can write $T^{\mu}{}_\nu = \lambda^4(\Lambda_0 + V_{vac}) \delta^\mu{}_\nu + \tau^\mu{}_\nu$, and so 
$T^\mu{}_\nu-\frac{1}{4} \delta^\mu{}_\nu \langle T^\alpha{}_\alpha \rangle = \tau^\mu{}_\nu-\frac{1}{4} \delta^\mu{}_\nu \langle \tau^\alpha{}_\alpha \rangle$: $\Lambda_0 + V_{vac}$ completely dropped out from the source in (\ref{eeqs}). 
There remains a `leftover' cosmological constant: the historic average  $\langle \tau^\mu{}_\mu \rangle/4$ contributes
to the curvature of the universe, {\it but without the classical and vacuum loop contributions}. Therefore we can write
\be 
M_{Pl}^2 G^\mu{}_\nu=  \tau^\mu{}_\nu-\frac{1}{4} \delta^\mu{}_\nu \langle  \tau^\alpha{}_\alpha \rangle \, ,
\label{eeqs1}
\ee
setting the sum of the classical Lagrangian and its quantum corrections to zero,  and forgetting them in what follows, at least in the limit
of (semi) classical gravity.

This is consistent since our action (\ref{action}) has {\it two} approximate symmetries which ensure the cancellations of the vacuum energy and protect the curvature from both large classical and quantum corrections \cite{wilczek,wein}. 
The first is the scaling $\lambda \rightarrow \Omega \lambda$, 
$g_{\mu\nu} \rightarrow \Omega^{-2} g_{\mu\nu}$ and $\Lambda \rightarrow \Omega^4 \Lambda$, broken only by the 
gravitational sector. The second involves the 
shift of $\Lambda$ and ${\cal L}$ in (\ref{action}) by $\alpha \lambda^4$ and $-\alpha$, so the action only changes by 
$\delta S = \sigma\left(\frac{\Lambda}{\lambda^4 \mu^4} + \frac{\alpha}{\mu^4}\right) - \sigma\left(\frac{\Lambda}{\lambda^4 \mu^4}\right)
\simeq \sigma' \frac{\alpha}{\mu^4}$. The scaling ensures that the vacuum energy at arbitrary order in the loop expansion couples to gravitational sector exactly the same way as the classical piece. The `shift symmetry' of the bulk action then cancels the matter vacuum energy and its quantum corrections\footnote{A similar behavior was observed in a different approach using historic integrals in \cite{linde2}.}. 
The scaling is broken by the gravitational action, but the breaking is mediated to the matter by the cosmological evolution,
through the scale dependence on $\int d^4 x \sqrt{g}$, and so is weak.
%, generating the mass gap in the matter sector. 
The residual 
cosmological constant is {\it small}: substituting the first of Eqs. (\ref{varsl}) and using $\lambda \propto m_{phys}/M_{Pl}$, we see that $\delta S \simeq \alpha \lambda^4 \int d^4 x \sqrt{g} \propto \alpha \left(\frac{m_{phys}}{M_{Pl}}\right)^4$, and is small when $m_{phys}/M_{Pl} \ll 1$, vanishing in the conformal limit\footnote{Fix $\int d^4 x \sqrt{g}$ and take $\mu \rightarrow \infty$ in the first of Eqs. (\ref{varsl}).} $\lambda \propto m_{phys} \rightarrow 0$. So, the bulk `shift symmetry' and the approximate scaling symmetry render 
a small residual curvature technically natural.

Quantum corrections from the matter sector to the Planck scale can be 
estimated by canonically normalizing ${\cal L}$ in (\ref{action}), and performing one loop renormalization of the Einstein-Hilbert Lagrangian. 
The corrections to $M_{Pl}$  from each 
species in the loop are given by \cite{myers}  $\Delta M^2_{Pl} \simeq {\cal O}(1) 
\times ({M}_{UV}^{phys})^2 + {\cal O}(1) \times 
m_{phys}^2 \ln({M}_{UV}^{phys}/m_{phys}) + {\cal O}(1) \times
m_{phys}^2 + \ldots$, where ${M}_{UV}^{phys}=\lambda {M}_{UV} $ is the matter UV regulator 
mass and $m_{phys}$ the mass of the virtual particle in the loop.
Thus, the Planck scale is radiatively stable\footnote{
The corrections alter our action (\ref{action}) qualitatively, changing
$M^2_{Pl} \rightarrow M^2_{Pl} +({M}_{UV}^{phys})^2= M^2_{Pl} +\lambda^2 {M}_{UV}^2$. This does not spoil the sequestering of vacuum energy in the protected sector: it
adds a term $- \frac{\lambda^2 { M_{UV}^2 }}{4} \langle R \rangle \delta^\mu{}_\nu$ to the rhs of Eq. (\ref{eeqs}). However 
this {\it vanishes identically} on shell, as long as $M_{Pl}^2 \ne 0$ \cite{kalpad2}.}
as long as ${M}_{UV}^{phys} \le M_{Pl}$, which is easily achieved in a sufficiently large and old Universe.
This is in contrast to the model discussed in \cite{tseytlin}, which does share some 
%nk superficial 
similarities with our mechanism. Indeed, imagine that instead of action (\ref{action}), we started with a theory 
$S = \int d^4x \sqrt{g} \left[\frac{\lambda^4 M_{Pl}^2}{2} R - \Lambda - \lambda^4 {\cal L}(g^{\mu\nu}, \Phi) \right] + \frac{\Lambda}{\lambda^4 \mu^4}$,
where we have chosen  a linear function $\sigma(z)=z$, and added a scaling with $\lambda$ in the Einstein-Hilbert term, 
but removed it from the matter Lagrangian.
We can readily integrate out $\Lambda, \lambda$, using $\lambda^4 = (\mu^4 \int d^4x \sqrt{g})^{-1}$ and 
$\frac{\Lambda}{\lambda^4 \mu^4} =  \int d^4x \sqrt{g} \left[\frac{\lambda^4 M_{Pl}^2}{2} R - \lambda^4 {\cal L}(g^{\mu\nu}, \Phi) \right] $, 
so that 
\be
S_{eff} = \frac{\int d^4x \sqrt{g} \left[\frac{M_{Pl}^2}{2} R - {\cal L}(g^{\mu\nu}, \Phi) \right]}{\mu^4 \int d^4x \sqrt{g}} \, .
\label{tseyt}
\ee
Although the variation removes the tree-level part of the cosmological constant \cite{tseytlin}, the radiative corrections survive. 
After conformally rescaling the metric  in (\ref{tseyt}) so that $M_{Pl}$ is independent of $\lambda$,  we see that the $\Lambda$ term scales as $\sim 1/\lambda^4$, and the  physical masses  as
$m_{phys}\simeq {m}/{\lambda^2}$. This implies that the radiative corrections to vacuum energy scale as $\sim 1/\lambda^8$, which differs from the scaling of the tree-level part, $\Lambda \sim 1/\lambda^4$. 
It was also noted that the theory (\ref{tseyt}) has Planck scale radiative instabilities. 
This stems from $\lambda^{4} = (\mu^4 \int d^4x \sqrt{g})^{-1}$ being small
in big and old universes, which makes the matter UV regulator mass
and the matter physical masses large 
(they scale like $\sim 1/\lambda^2$)
relative to $M_{Pl}$, so that $M_{Pl}$ is susceptible to the 
renormalization effects from them.
None of this is a problem for our mechanism in (\ref{action}).

Let us consider now our historic average, 
$\langle \tau^\alpha{}_\alpha \rangle$. In our case, the individual factors in the ratio must be finite too. First, $\int d^4 x\sqrt{g}$ must be finite: 
{\it i}) we require $\sigma(z)$ to be differentiable, to get field equations (\ref{eeqs1}); {\it ii}) hence, divergent 
$\int d^4x \sqrt{g}$ would generically force $\lambda$ to vanish; {\it iii)} but $\lambda \ne 0$ since $m_{phys} \propto \lambda$ in the matter
sector. Fortunately there is a diffeomorhism invariant regulator for these integrals: {\it spacetime singularities}. A spatially compact universe of finite lifetime, starting in a bang and ending with a crunch, 
has finite integral $\int d^4x \sqrt{g} = {\cal O}(1) {\rm Vol}_3/H_{age}^4$, where ${\rm Vol}_3$ is the comoving spatial volume, and 
$H^{-1}_{age}$ is the scale of the lifetime of the universe. 
Furthermore, for sources which 
obey the standard energy conditions ($|p/\rho | \le 1$), we can estimate \cite{kalpad2} $\int d^4 x \sqrt{g} \, \tau^\mu{}_\mu \sim - {\rm Vol}_3 \int_{t_{bang}}^{t_{crunch}} dt a^3 \rho$, in comoving coordinates. The only potentially divergent contributions  come from the end points, where $\rho$ scales as $\rho \sim 1/(t-t_{end})^2$, by virtue of the Friedman equation, where $t_{end}$ is either of the instants of bang or crunch. In this limit , $a^3 \sim (t-t_{end})^{2/(1+w)}$, and so the integrand is $a^3 \rho \sim (t-t_{end})^{-2w/(1+w)}$. The integral will not diverge 
provided $|w| \le 1$\footnote{If $w=+1$, these contributions will diverge at most logarithmically, with ${\cal O}(M_{Pl}^2 H_0^2)$ coefficients. When properly cut off at the physical singularities $t_{Pl} = M_{Pl}^{-1}$ they will be finite, and much smaller than the cutoff.}.
So, for realistic matter sources,  our historic averages will always be finite in a bang/crunch universe. 
Next, it is straightforward to show \cite{kalpad2} that the largest contribution to $\langle \tau^\mu{}_\mu \rangle$ will come from the turnaround region, when the Universe is close to its maximal size. We then find that 
$\langle \tau^\mu{}_\mu \rangle \simeq {\cal O}(1) M_{Pl}^2 H_{age}^2$, where we recall that the scale of the lifetime of the universe, $H^{-1}_{age}> H_0^{-1}$, where $H_0^{-1}$ is its current age. This would yield a naturally small cosmological constant in our universe 
(with the sign controlled by the pressure of the dominant contribution) if it begins to collapse in, say, 
100 billion years or so. 
%nk This might happen if our universe were either spatially closed, with a small but nonzero 
%positive spatial curvature, or the current acceleration were a transient, with the net potential 
%turning negative some time in the future. 
This might happen if  the current acceleration were a transient, with the net potential turning negative some time in the future, and/or our universe were spatially closed, with a small but nonzero positive spatial curvature. 
%nk
This might not be impossible. For example, the current LHC data suggest that the Higgs potential may indeed have an unstable phase, with the Higgs 
{\it vev} close to the precipice \cite{gian}. 
%nk More 
Curiously, a warning about this has been raised in the prescient paper by Wilczek 
quite a while ago \cite{wilczek}. 

What about the contributions to the cosmological constant from phase transitions in the early universe \cite{dreitlein,linde,veltman}? 
In our setup they do {\it not} drop out from (\ref{eeqs},\ref{eeqs1}), but they become {\it automatically} small
at times after the transition in a large and old universe. 
To see it, we model them with a step function\footnote{This ignores gradient corrections, but  the error
is at most ${\cal O}(1)$. } potential   $V = V_{before} (1- \Theta(t-t_{*}))+ V_{after} \Theta(t-t_{*})$, where $ \Theta(t-t_{*})$ is the step function, and $t_{*}$ the transition time. Substituting into  (\ref{eeqs1}), {\it after} the transition we find
\be
\tau^\mu{}_\nu-\frac{1}{4} \delta^\mu{}_\nu \langle  \tau^\alpha{}_\alpha \rangle \simeq \delta^\mu{}_\nu \frac{\int d^4x \sqrt{g} (V-V_{after})}{\int d^4x \sqrt{g}} 
 \simeq  \delta^\mu{}_\nu \left(\frac{\Delta V}{M_{Pl}^2 H_*^2} \right) M_{Pl}^2 H_{0}^2 \left(\frac{H_{age}}{H_0} \right)^2 \left(\frac{H_{age}}{H_*} \right)^\frac{1-w}{1+w} 
\label{phase}
\ee
where  $\Delta V = V_{before} - V_{after}$, and $H_*$ is the curvature scale during the transition, of the order of $\sqrt{V_{before}}/M_{Pl} \ga \sqrt{\Delta V}/M_{Pl}$. For simplicity, we took the matter from the transition to turnaround to be a single component fluid with a fixed $w$; a more precise estimate would merely give corrections of order one, provided we restrict attention to physically reasonable matter sources with $|p/\rho|\le1$ 
\cite{kalpad2}. In any event, as long as $H_* \gg H_{age}$, which is true for the Standard Model, the vacuum energy contributions from early phase transitions are far smaller than the current critical density $M_{Pl}^2 H_0^2$.  

How could a universe become so big in our framework? The simplest mechanism to explain it is inflation. To incorporate it in the theory, we can 
add an extra sector to (\ref{action}) which contains an inflaton, outside of the protected sector ${\cal L}$. A slightly nontrivial issue is that once inflation ends, the universe needs to reheat by particle production in the protected sector, so the 
%nk 
inflaton must couple to the fields given in ${\cal L}$. A model which realizes this without spoiling the sequestration of vacuum energy from ${\cal L}$ is the original inflation of Starobinsky \cite{star}, which is actually the model preferred by the current data anyway \cite{data}. So we just add a term 
$\int d^4x \sqrt{g} \, \beta \, R^2$ to the action (\ref{action}) where $\beta \sim {\cal O}(10^6)$ is a dimensionless parameter. 
This is radiatively stable under the protected sector loops due to $\beta$ being so large\footnote{One can easily see that from \cite{myers}; further, the $\lambda$-dependence cancels as claimed once we pick Pauli-Villars regulators that couple to $\tilde g_{\mu\nu}$ in (\ref{action})}. In line with our philosophy here, we will treat this term as a semi-classical term in the theory, still ignoring any loops with virtual gravitons. In the axial gauge, extracting the Starobinsky scalaron $\chi$ by the field redefinition $\bar g_{\mu\nu} = \left(1+\frac{4\beta}{M_{Pl}^2} R\right) g_{\mu\nu}$, 
$\chi = \sqrt{\frac32} M_{Pl} \ln\left(1+\frac{4\beta}{M_{Pl}^2} R\right)$ \cite{kko}, we treat $\chi$ as a (semi) classical field too, omitting any processes where it appears in loops. The scalaron has the potential $V_\chi = \frac{M_{Pl}^4}{16 \beta} \left[1- \exp\left(-\sqrt{\frac23} \chi/M_{Pl}\right)\right]^2$ and the
matter couples to both $\bar g_{\mu\nu}$ and to $\chi$, via
\ba
S &=&   \int d^4 x \sqrt{\bar g} \Bigl[ \frac{M^2_{Pl}}{2} \bar R  - \frac12 (\bar \partial \chi)^2 - V_\chi - \Lambda e^{-2\sqrt{\frac23} \frac{\chi}{M_{Pl}}}  -
{\lambda^4} e^{-2\sqrt{\frac23} \frac{\chi}{M_{Pl}}} {\cal L}(\lambda^{-2} e^{\sqrt{\frac23} \frac{\chi}{M_{Pl}}} \bar g^{\mu\nu} , \Phi) \Bigr]  \nonumber \\
&& ~~~~~~~~~~~~  +
\sigma\left(\frac{ \Lambda}{\lambda^4 \mu^4}\right) \, . 
\label{star}
\ea
The dynamics of inflation and reheating is almost the same as in the Starobinsky model. The only difference is that now 
$\frac{\sigma'}{\lambda^4\mu^4}  = \int d^4x \sqrt{g} e^{-2\sqrt{\frac23} \frac{\chi}{M_{Pl}}}$, 
which involves $\chi$, only shifts
the numerical value of $\lambda$ by 
%nk a 
very little. This is because $\chi \ne 0$ only during inflation, while the dominant contribution
comes from the full history of the universe. 

As we noted above, the parameter $\lambda$ controls the physical scales in ${\cal L}$, setting $m_{phys} = \lambda m$. It cannot protect the hierarchy between $m_{phys}$ and $M_{Pl}$, and the hierarchies between different physical masses in ${\cal L}$. But it can help set it, coexisting with models which address particle hierarchies and help them solve the vacuum energy problem. As an example, the regulator of the
protected sector in ${\cal L}$ can be as high as $M_{UV}^{phys} \sim  M_{Pl}$. This 
requires $\lambda \sim {\cal O}(1)$, and may imply a vacuum energy as high as $\Lambda \sim M_{Pl}^4$, which is nevertheless sequestered 
from gravity by our mechanism. If we take the universe to have a lifetime ${\cal O}(10) H_0^{-1}$, the first of Eqs. (\ref{varsl}) implies 
$\sigma' \sim (10  \mu/H_0)^4$. Taking $\mu \sim 0.1 M_{Pl}$ and $\sigma \simeq e^{\frac{\Lambda}{\lambda^4 \mu^4}}$ we can 
account for such a large vacuum energy\footnote{To allow for either sign of $\Lambda$ for 
a fixed $\lambda > 0$, take eg. $\sigma = \sinh\left(\frac{\Lambda}{\lambda^4 \mu^4}\right)$.}. The Standard Model may be embedded in ${\cal L}$ via some of its BSM extensions, such as eg. some variant of supersymmetry, in which case we could get a much lower value of $\Lambda$, given by the fourth power of the SUSY breaking scale. If this is as low as TeV, then we require  $\mu \sim$ TeV. Either way, $\mu$ can be chosen to fit whatever mechanism protects the hierarchy within ${\cal L}$ so that our proposal can then be utilized to sequester the vacuum energy contributions to ${\cal L}$ which the BSM extension cannot remove.

Why can our mechanism sequester the vacuum energy of the protected sector, both classical and quantum?
The problem, as explained in the context of the Weinberg's no-go theorem \cite{wein}, is with the trace equation, which involves $\Lambda$ as an {\it a priori} arbitrary source. 
Because $g$ is a pure gauge mode, its variational equation does not provide any intrinsic boundary conditions - all are equally good, by symmetry. 
The integral $\int d^4 x \sqrt{g}$ is gauge (i.e., diffeomorphism) invariant, but 
it is not independent: its variation is a linear combination of the variation of $g_{\mu\nu}$ at all spacetime points. So since $\int d^4x \sqrt{g}$ multiplies 
the vacuum energy, its variation yields a source $\propto \Lambda$ as opposed to constraining it to vanish or to be small.
Our mechanism  dramatically changes the role of $\int d^4x \sqrt{g}$.  Since it is a true scalar, we make all the physical scales in the protected matter sector depend on it, which automatically forces the vacuum energy to drop out. As an example, if $\sigma(z)=z$ in (\ref{action}) 
and ${\cal L}$ is literally the Standard Model,
we can integrate $\Lambda$ and $\lambda$ out by using (\ref{varsl}) and rewrite (\ref{action}) as just Einstein-Hilbert action coupled 
to the Standard Model, with the only modification being that the Higgs vev $v$ is replaced by $v/(\mu^4 \int d^4 x \sqrt{g})^{1/4}$.
Further, in (asymptotically) flat space, the integral $\int d^4x \sqrt{g}$ is infinite, which would send the physical matter scales to zero, yielding the same outcome as in GR, as sanctioned by Weinberg's no-go theorem. In a collapsing spacetime however
$\int d^4x \sqrt{g}$ is finite, gapping the particle spectrum from zero, mediating cosmologically the scaling symmetry breaking in the gravitational sector (giving a residual cosmological constant 
$\langle \tau^\mu{}_\mu \rangle/4$,  which is, however, completely independent of the cutoff and naturally small in a large old universe by virtue of the two approximate symmetries). This scale dependence on $\int d^4x \sqrt{g}$ is completely invisible to any nongravitational local experiment, by diffeomorphism invariance. Since no new propagating modes appear, locally the theory looks just like standard GR, in (semi) classical limit {\it but without a large cosmological constant}. 

Cosmic eschatology changes, however, since consistency requires that a universe should have a compact spacetime, whose signatures could be sought for in cosmology, both in the frozen sky and in its evolution. 
The mechanism also predicts that there should be a residual cosmological constant, which is automatically small in an old and big universe, and it would be interesting to seek for the right ingredients that could make it fit the current data. 
Since the universe eventually collapses, the residual cosmological constant cannot dominate forever, and so 
$w_{DE} \simeq -1$ as determined from the data is a (possibly long lived) transient state. 
One expects that there would be significant 
differences in the regime of quantum gravity, however, but that's hardly constrained by any currently known facts. 
Clearly these issues should be explored further, and we intend to address some in more detail in \cite{kalpad2}.

\vskip.2cm

{\bf Acknowledgments}: 
We would like to thank Guido D'Amico, Savas Dimopoulos, Matt Kleban, Albion Lawrence, Fernando Quevedo and Paul Saffin for very useful discussions.
NK thanks the School of Physics and Astronomy, U. of Nottingham for hospitality in the course of this work.
NK is supported by the DOE Grant DE-FG03-91ER40674, and by a Leverhulme visiting professorship.
AP was funded by a Royal Society URF.

\end{document}